
%
%
\magnification=\magstep1
\vsize8.4truein
\font\smallfont=cmr8                 
\abovedisplayskip=18pt plus 5pt minus 13pt  
\belowdisplayskip=18pt plus 4pt minus 13pt  
\def\topglue{                               
  \nointerlineskip \vglue-\topskip \vglue%
}
\def\refitem{\par\noindent\hangindent=30pt\hangafter=1}
%
%
\def\lroughly{                              
  {<\kern-1.1em \lower.9ex\hbox{$\sim$}\kern.5em}
}
\def\groughly{                              
  {>\kern-1.1em \lower.9ex\hbox{$\sim$}\kern.5em}
}
\def\hone{${}^1$H}                          
\def\htwo{${}^2$H}                          
\def\hthree{${}^3$H}
\def\hethree{${}^3$He}
\def\hefour{${}^4$He}
\def\lisix{${}^6$Li}
\def\liseven{${}^7$Li}
\def\lieight{${}^8$Li}
\def\linine{${}^9$Li}

\def\benine{${}^9$Be}
\def\bten{${}^{10}$B}
\def\beleven{${}^{11}$B}
\def\celeven{${}^{11}$C}
\def\csixteen{${}^{16}$C}
\def\nfourteen{${}^{14}$N}
\def\nfifteen{${}^{15}$N}
\def\nsixteen{${}^{16}$N}
\def\nseventeen{${}^{17}$N}
\def\ofifteen{${}^{15}$O}
\def\oeighteen{${}^{18}$O}
\def\onineteen{${}^{19}$O}
\def\otwenty{${}^{20}$O}
\def\fnineteen{${}^{19}$F}
\def\ftwenty{${}^{20}$F}
\def\ftwentyone{${}^{21}$F}
\def\netwentyone{${}^{21}$Ne}
\def\e{{\hbox{e}}}
\def\n{{\hbox{n}}}
\def\p{{\hbox{p}}}
\def\d{{\hbox{d}}}
\def\t{{\hbox{t}}}
\def\h{{\hbox{H}}}
\def\He{{\hbox{He}}}
\def\Li{{\hbox{Li}}}
\def\Be{{\hbox{Be}}}
\def\B{{\hbox{B}}}
\def\C{{\hbox{C}}}
\def\N{{\hbox{N}}}
\def\O{{\hbox{O}}}
\def\F{{\hbox{F}}}
\def\Ne{{\hbox{Ne}}}
\def\rarr{\rightarrow}

%

%
%
%
%

\hfill UMN-TH-1020/92

\hfill June 1992

\topglue 1in
\centerline{\bf Primordial Nucleosynthesis and the Abundances of
                Beryllium and Boron}

\vskip1cm
\centerline{David Thomas${}^1$, David N.~Schramm${}^{1,2}$,
            Keith A.~Olive${}^3$, Brian D.~Fields${}^1$}

\vskip0.4cm
\centerline{${}^1$ University of Chicago, Chicago, IL 60637}
\centerline{${}^2$ NASA/Fermilab Astrophysics Center, FNAL, Box 500
            Batavia, IL 60512}
\centerline{${}^3$ University of Minnesota,
            School of Physics and Astronomy, Minneapolis, MN 55455}

\vskip2cm
\centerline{Abstract.}

\midinsert\narrower\narrower\noindent
The ability to now make measurements of Be and B as well as put
constraints on \lisix\ abundances in metal-poor stars has led to a
detailed reexamination of Big Bang Nucleosynthesis in the
$A\groughly6$ regime.  The nuclear reaction network has been
significantly expanded with many new rates added.  It is demonstrated
that although a number of $A>7$ reaction rates are poorly determined,
even with extreme values chosen, the standard homogeneous model is
unable to produce significant yields (Be/H and B/H $<10^{-17}$ when
$A\le7$ abundances fit) above $A=7$ and the \liseven/\lisix\ ratio
always exceeds 500.  We also preliminarily explore inhomogeneous
models, such as those inspired by a first order quark-hadron phase
transition, where regions with high neutron/proton ratios can allow
some leakage up to $A>7$.  However models that fit the $A\le7$
abundances still seem to have difficulty in obtaining significant
$A>7$ yields.
\endinsert

\vfil\eject
\baselineskip=20pt 

\centerline {1. Introduction}

Over the last quarter century the standard homogeneous model of Big
Bang nucleosynthesis has proved spectacularly successful at predicting
the primordial abundances of the light elements.  In particular,
Homogeneous Big Bang Nucleosynthesis (BBN) (see Walker et al.\  1991 and
references therein) successfully fits \hone, \htwo, \hthree,
\hefour, and \liseven\ abundances in primordial objects (extremely
low $Z$) over a dynamic range in abundance of almost ten orders of
magnitude.  The fit to these abundances has become the prime
determinator of the cosmological density of baryons
($\Omega_b\sim0.05$ where $\Omega_b$ is the fraction of the critical
density in baryons).  Also of significance has been BBN's prediction
of the number of neutrino flavors (Steigman, Schramm \& Gunn 1977,
Yang et al.\ 1984, Walker et al.\ 1991) which has now been confirmed in
accelerator experiments.

Because of the double particle instability gaps at $A=5$ and $A=8$ it
was rapidly recognized that homogeneous BBN produces small amounts of
$A=7$ (\liseven/H$\sim10^{-10}$) and no significant yields for $A>7$.
Recently observers have begun to be able to observe Be and B in
extreme pop.\ II stars with low $Z$ (Rebolo et al.\ 1988, Ryan et
al.\ 1990, Gilmore, Edvardsson \& Nissen 1991, Duncan, Lambert \& Lemke
1992, Gilmore et al.\ 1992, Ryan et al.\ 1992) and have begun to put
limits on \lisix\ in similar objects (Andersen, Gustafsson \& Lambert
1984, Spite \& Spite 1982, Hobbs \& Pilachowski 1988, Hobbs \&
Thornburn 1991).  Data to date on Be and B are best understood as
being due to cosmic ray spallation (Walker et al.\ 1992, Steigman \&
Walker 1992) and \lisix\ is still undetected.  However, because of the
potential for new developments here, we have reexamined BBN yields
with a particular focus on $A\ge6$.  While the basic conclusions of
the earlier calculations remain unchanged, we did note that the
networks used earlier were not particularly complete for $A\ge6$.
(This relative incompleteness had little or no effect on $A<6$
yields.)  Therefore we have extended the nuclear reaction network and
included many links that were missing (or poorly estimated) in earlier
calculations.  Where links in our new network are poorly known we have
run the calculation with a range of values to assess the sensitvity to
those links.  We feel this present paper is the most thorough
exploration of high $A(\ge6)$ Homogeneous Big Bang Nucleosynthesis
done to date.

A much discussed alternative to Homogeneous Big Bang Nucleosynthesis
has been the first-order quark-hadron phase-transition inspired
inhomogeneous model (Alcock, Fuller \& Mathews 1987, Applegate, Hogan
\& Scherrer 1988, Turner 1988, Terasawa \& Sato 1989, Kurki-Suonio et
al.\ 1990).  It had been proposed (Boyd \& Kajino 1989, Malaney \& Fowler
1989) that in such a model, the high n/p regions may allow leakage
beyond the $A=8$ gap and produce interesting amounts of Be and B.
While Terasawa \& Sato (1990) argued against such leakage, we noted that
their network was not as complete as the one we have developed for the
homogeneous case and that our more complete network may be needed to
fully explore the situation.  In addition, with our enhanced network,
we are able to quantitatively compare the yields of the heavier
elements in the homogeneous and inhomogeneous models.

Thus in this paper we will also do a preliminary exploration of high
n/p conditions with a homogeneous calculation.  (Since high n/p values
only occur in the low density zones of inhomogeneous models, high
baryon to photon results are less significant than the low baryon to
photon results for our high n/p calculation.)  In a future paper we
will do a more complete exploration of inhomogeneous models using a
full multizoned model with back diffusion as in Alcock, Fuller \&
Mathews (1987), Kurki-Suonio et al.\ (1990), Terasawa \& Sato (1990).
However, at present to get a feel for the full network and see the
critical points in the calculation we felt this simpler exploration
was necessary and illustrative of the network itself.

One point we emphasize is that the high $A$ calculations cannot be
used without making sure that the low $A$ abundances are fit.  As
Kurki-Suonio et al noted, the inhomogeneous calculations still require
about the same $\Omega_b$ as the standard homogeneous BBN model.  Thus
any leakage up to high $A$ must occur with standard $\Omega_b$ values
if it is to be relevant.

\bigskip
\centerline{2. The Reaction Network}

Up until now, the emphasis in measuring astrophysically relevant
nuclear reaction rates has generally been on those reactions involved
in the production of elements up to Li.  Of the yields for $A\le7$
only \liseven\ has been found to be sensitive (Kawano et al.\ 1988,
Krauss \& Romanelli.\ 1990) to uncertainties in the measured reaction
rates.

We have updated the reaction network in an attempt to resolve this
question for $A\ge6$.  In table 1 we list the reactions that have been
added (A) or updated (U) to the last version of the code (Walker et
al.\ 1991) which itself is an updated version of Wagoner's code (Wagoner
1969), modified to evaluate neutron--proton weak interaction rates
according to the prescription of Walker et al.\ (1991).  The majority of
the reaction rates used came from Caughlan and Fowler (1988).  To this
set we added reactions from Wagoner (1969), Lederer \& Shirley (1978),
Endt \& Van der Leun (1978), Ajzenberg-Selove (1983),
Tuli (1985), Malaney \& Fowler (1989), Boyd \& Kajino (1989),
Wiescher, Steininger \& K\"appeler (1989), Wang, Vogelaar \& Kavanagh
(1991), Kawano et al.\ (1991), Brune, Kavanagh,
Kellogg \& Wang (1991), Barhoumi et al.\ (1991), Becchetti et al.\ (1992),
Boyd et al.\ (1992a), and Boyd et al.\ (1992b).  We also included reactions
from Smith, Kawano \& Malaney (1992) as given in Kawano (1992).
In a number of cases
(listed in table 1) we were unable to find any reliable figures for
reactions in the relevant temperature range and made estimates (E).
For a few reactions we have found more than one rate published
recently and have tested for sensitivity (S) when these differed
beyond stated errors.  The complete network is shown in fig.\ 1.  For
the most part we used the most recently published rate, whenever more
than one expression was available.  For the reactions labelled S in
table 1 we ran the code using the rates from each of the sources
listed.  For estimated rates or rates where experimental values
differed outside of the published error bars we made tests of the
sensitivity of our results by arbitrarily increasing and decreasing
that particular rate by a factor of 1000 or by using the extreme
experimental value to see the sensitivity of the result.  The actual
prefered reaction rates selected are shown in table 1, and the explicit
rates used are given in Appendix 1.

In order to ensure that our network is sufficiently extensive, we have
plotted flow diagrams (see figs.\ 2) for $\eta_{10}=1.0$ for both the
standard calculation, and for high n/p
($\eta_{10}=\eta\times10^{10}$).  These were produced by connecting
the nuclide with the greatest increase in mass fraction, to that with
the greatest decrease (omitting nuclides with $A\leq4$) at each time
step.  This gives a pictorial representation of which links are most
significant in producing any given nuclide.  Fig.\ 2a shows that in
the standard model, most of the flow proceeds along the central
portion of the network, with the exception of the flow to \oeighteen.
It was this \oeighteen\ that prompted us to add \csixteen, \nsixteen,
\nseventeen, \onineteen, \otwenty, \fnineteen, \ftwenty,
\ftwentyone\ and \netwentyone\  to our original network, however
this made no difference to the yields of the light elements.  Fig.\ 2b
shows that for $n/p=10$, even with the extended network, there is a
significant flow on the neutron-rich side, particularly to \oeighteen\
and \onineteen.  Obviously, to accurately explore neutron-rich flows
requires networks as rich as the one we use here.  Previous explorations
of $A>7$ yields have not used such an extensive network.

\bigskip
\centerline{3. The Results}

The yields of the light elements are shown as functions of the baryon
to photon ratio $\eta$, for a neutron mean-life of 889.6 sec.\ in
figs.\ 3.  Figure 3a shows the \hefour\ mass fraction as a function of
the baryon to photon ratio, $\eta=n_B/n_\gamma$ and figure 3b shows
the abundances of \htwo, \hethree, \lisix\ and \liseven.  In
particular, \liseven\ is always greater than \lisix\ by a factor of at
least 500 (at $\eta_{10} = 0.01$), while for $\eta_{10}\sim3$ (where
calculations agree with observations) the \liseven/\lisix\ ratio is
$4700$.

Of the reactions labelled S in table 1, the only rate whose variation
has a significant effect on the results is \liseven(t,n)\benine.  This
sensitivity is shown in fig.\ 3c.  Furthermore, variation in this
reaction affects only \benine\ and \bten.  Yields of the lighter
elements (H, He, Li) are unaffected by the variation in any of the
rates S.

The double hump for \beleven\ is a result of the fact that it can be
created directly (for low $\eta$) or through \celeven\ (high $\eta$)
which then $\beta$-decays to \beleven.  The highest yield for both
\benine\ and \bten\ is given by the \liseven(t,n)\benine\ rate taken from
Boyd \& Kajino (1989).  The lowest is from Malaney \& Fowler (1989).
Note that in no case is it possible to produce a \benine\ number density
relative to hydrogen greater than about $10^{-14}$.  If we consider
the observational limits on H, He and Li then $\eta$ is constrained by
$2.8\lroughly\eta_{10}\lroughly4.0$ (Walker et al.\ 1991).  In this
range \benine/H has a maximum yield of $6\times10^{-18}$, and
\beleven/H a maximum of $8\times10^{-17}$.

For the reactions labelled E it was necessary to make estimates of the
rates.  Unfortunately the presence of resonances at nucleosynthesis
energies can influence the reaction rate by many orders of magnitude.
We have tried a number of approaches to this problem.  Some of the reactions
in question are given in Delano (1969). However, Delano parameterizes his
rates for $1\leq T_9\leq10$ and
is not explicit about his estimation methods.  Nor are they
transparent from his rate expressions which are all fit to a standard
form.  The only theory he mentions is a statistical nuclear model which is
only accurate for heavy nuclei with a large, even distribution of levels
lying within nucleosynthesis energies.

To understand the possible effects of the unmeasured rates, one would ideally
like to at least have upper bounds on them.  Such limits are enormous,
however, as resonances at the effective nucleosynthesis energy
$$
E_{\hbox{\smallfont eff}}
    = [\pi \alpha Z_0 Z_1 kT (\mu c^2/2)^{1/2}]^{2/3}
    = 0.12204 Z_0^{2/3} Z_1^{2/3}
    {{A_0 A_1}\over{A_0+A_1}} T_9^{2/3} \, {\hbox{MeV}}\eqno{(1)}
$$
(where $\alpha \approx 1/137$, $Z_0$ and $Z_1$ are the charges of the
incoming nuclei, and $\mu$ their reduced mass) can enhance a reaction
rate to many
orders of magnitude larger than reactions that are nonresonant, or occur
through the tail of a resonance. We have made estimates of the rates to
compare with Delano's.  To do this we assumed each reaction had
two contributions, one nonresonant and one resonant.  In both cases the
reaction rates can be given by expressions of the form
$$
\langle \sigma v \rangle_T = \lambda_0 T^{-n} \exp( - a/T^m )\eqno{(2)}
$$
where the constants $\lambda_0$, $a$, $n$ and $m$ differ for the two
cases.

In the nonresonant case, the exponential part of the rate is
determined by the nuclear masses and charges, so that it is only the
preexponential factor which is unknown.  We have estimated this factor
(the unknown
part of which is the cross section factor $S(0)$) using the
method of Fowler \& Hoyle (1964), and making
conservative estimates of optical model parameters we have
$$
\langle\bar\sigma v\rangle \approx\left({2\over\mu}\right)^{5/6}
\left({2\hbar\over kT}\right)^{2/3}
{(2\pi e^2Z_0Z_1)^{4/3}\over(3V_0)^{1/2}}\exp(2x-\tau) \eqno{(3)}
$$
where $V_0\approx40$ MeV gives a good fit to known cross-sections,
$$
x=2\sqrt{2\mu Z_0Z_1e^2R/\hbar^2};\quad R=R_0(A_0^{1/3}+A_1^{1/3});\quad
R_0=1.44\hbox{fm} \eqno{(4)}
$$
and
$$
\tau=3E_{\hbox{\smallfont eff}}/kT \eqno{(5)}
$$
  Thus our rate
is computed
in an appropriate low-energy limit, but arises from an optical model which
assumes a smoothly varying and large level density which we do not expect
to be very accurate for the light nuclei we consider.

In the resonant case, we used a recent tabulation by Ajzenberg-Selove
(1985, 1986, 1988)
to locate levels and to find their widths.  Knowing the resonant energy fixes
the exponential term in the rate, and so again we are uncertain only in the
preexponential factor.  The unknown part here is the reduced width
$(\omega \gamma)_r \sim \sigma_r E_r \Gamma_r$, in which the cross section
at resonance, $\sigma_r$ is unknown.  For this we have assumed an arbitrary
but generous value of 1 barn.

Using these estimates can have a significant effect on nuclear
yields (although not for \benine\ or any of the lighter elements).
In particular, \bten($\alpha,\gamma$)\nfourteen\ and
\beleven($\alpha,\gamma$)\nfifteen\ reduce the yields of \bten\ and
\beleven\ respectively by two or three orders of magnitude at $\eta_{10}=3$,
and while \celeven($\alpha,\gamma$)\ofifteen\ has no effect at $\eta_{10}=3$,
it removes \celeven\ almost completely above $\eta_{10}=10$.  We emphasize
however, that these reaction rates are dominated by contributions from
resonances near the entrance channel, and that we
consider them to be extreme upper limits.

In order to find more reasonable limits we have also tried using Delano's
expressions directly (with results identical to those obtained by omitting
all reactions E) and estimating rates on the basis of ``similar'' reactions.
For each reaction E, we used a rate which was higher than the rate of any
reaction with a similar form (say all reactions
X(n,p)Y for the case of ${}^{12}$N(n,p)${}^{12}$C), and also higher than
Delano's rate (where one was available) within the range $1\leq T_9\leq10$,
and than increased the result by a factor of 1000.  The only reaction for
which this had any effect on the results was
${}^9$Li(d,n)${}^{10}$Be.  Using 1000 times the rate for
${}^6$Li(d,n)${}^7$Be (Malaney \& Fowler 1989) increases the yield of
\bten\ by up to a factor of 2 for $\eta_{10}\lroughly0.1$.  (For
$\eta_{10}\groughly0.1$ the effect is again insignificant.)

Clearly further experimental data would be helpful, however we believe
(with a few special exceptions that will be mentioned later) that
these reactions are likely to have little, if any, effect on the light element
abundances.

The significance of the Be and B calculations has been made evident by
recent observations in a number of low metallicity halo dwarf stars of
these two elements.  It is generally known that big bang
nucleosynthesis is incapable of producing an observable amount of
either of these two light elements.  In fact, Be and B have generally
been thought to have been produced by cosmic ray spallation.  Indeed,
in recent analyses, the observed Be and B have been argued to be
explicable entirely in terms of cosmic ray spallation in the early
galaxy, while maintaining consistency with big bang nucleosynthesis .
(Spallation also produces \liseven\ which thereby reduces the required
production from big bang nucleosynthesis, but as emphasized by Olive \&
Schramm (1992) this reduction is still completely compatible with
the other cosmological light element abundances).  Furthermore, it has
been speculated that inhomogeneous models may provide for enhanced Be
and B abundances relative to the standard model.  Here, we have found
that the standard model production of Be and B is indeed negligible
relative to the observations and in our exploration of high n/p we
find that while Be and B yields are enhanced, they are still well
below the observations.

\bigskip
\centerline{4. Limits on Inhomogeneous Yields}

To obtain an extreme upper limit on the yields produced by
inhomogeneous nucleosynthesis, we show in figs.\ 4a,b the results of
running the code with the initial neutron-proton ratio raised, for
$\eta_{10}=3.0$.  (In these runs we have frozen the n/p ratio at the
value given on the ordinate axis for temperatures down to $T_9=5$,
below which the calculation is allowed to proceed as normal.)  Raising
n/p has the effect of increasing the yields of the light elements,
however it is important to note that in an accurate calculation the
large yields in the high n/p regions are diluted by the smaller yields
in the low n/p regions as well as by interactions at the interface of
the regions.

Note that increasing n/p ceases to cause significant effects for
n/p$\groughly3$, and any realistic calculation with multizones
(e.g.\ Kurki-Suonio et al.\ 1990) has been found to have much back-diffusion,
reducing the magnitude of any extremes.  Fig.\ 4b shows that in our
extreme cases \benine\ saturates at a yield of $\lroughly10^{-12}$,
however, note that \hefour\ is overproduced in all high n/p zones by a
factor of $\sim4$.  Thus the minimum reduction must be at least a
factor of 4, and as mentioned before, realistic multizone models will
yield even greater reductions.  In a future paper we will investigate
more thoroughly the effects of our updated network on inhomogeneous
nucleosynthesis, however from the current preliminary exploration it
seems unlikely that the yields can be sufficient to produce the Be and
B abundances observed in some Pop II stars (Ryan et al 1990, Gilmore,
Edvardsson \& Nissen 1991, Duncan, Lambert \& Lemke 1992).

\bigskip

\centerline{5. Conclusions}

We have shown that even with our extended reaction network, the standard
homogeneous model of primordial nucleosynthesis is unable to produce
significant yields of Be and B.  In addition, it appears that inhomogeneous
nucleosynthesis is unlikely to produce much greater yields, even though
there are uncertainties in some reactions.

Finally, we feel it would be useful to have further data on those nuclear
reaction rates marked S and E in table 1.  In particular,
\liseven(t,n)\benine, \lieight(p,$\gamma$)\benine\ and
\linine(p,n)\benine\ may have a measurable effect on \benine\ production.
\bten($\alpha,\gamma$)\nfourteen, \beleven($\alpha,\gamma$)\nfifteen\ and
\celeven($\alpha,\gamma$)\ofifteen\ may also be critical in the production
of the heavier elements.

\bigskip

\centerline{Acknowledgements.}

We would like to thank Gary Steigman, Jim Truran and Terry Walker for
helpful discussions.  This work was supported in part by DoE grant
DE-AC02-83ER40105 at Minnesota, by the DoE, NASA and NSF at Chicago,
and by NASA through grant NAGW 2381 at Fermilab.

\vfill\eject

\centerline{Appendix 1---Reaction Rates}

We present below the rates, in Fortran notation, used for reactions
introduced or updated
since Walker et al (1991).  The notation is the usual one, used in
astrophysical reaction rate tabulations, in which
T9ab $=T_9^{a/b}$, T9Mab $=T_9^{-a/b}$ and the expressions represent
values for $F\equiv N_A\langle\sigma v\rangle$, where $N_A$ is Avogadro's
number, $\sigma$ the cross-section, $v$ relative velocity, and the
thermal average is taken over a Boltzmann distribution.


\smallskip\noindent${}^{3}\h+\e\rarr{}^{3}\He$\par  
\f@
      F = 1.78E-9
@
\smallskip\noindent
${}^{8}\Li\rarr\e+2{}^{4}\He$\par            
\f@
      F = 8.27E-1
@
\smallskip\noindent
${}^{11}\Be\rarr\e+{}^{11}\B$\par            
\f@
      F = 0.0502
@
\smallskip\noindent
${}^{8}\B+\e\rarr2{}^4\He$\par               
\f@
      F = 9.00E-1
@
\smallskip\noindent
${}^{12}\B\rarr\e+{}^{12}\C$\par             
\f@
      F = 3.43E+1
@
\smallskip\noindent
${}^{11}\C+\e\rarr{}^{11}\B$\par             
\f@
      F = 5.67E-4
@
\smallskip\noindent
${}^{14}\O+\e\rarr {}^{14}\N$\par            
\f@
      F = 9.82E-3
@
\smallskip\noindent
${}^{15}\O+\e\rarr {}^{15}\N$\par            
\f@
      F = 5.67E-3
@
\smallskip\noindent
${}^{17}\F+\e\rarr{}^{17}\O$\par             
\f@
      F  = 0.0107
@
\smallskip\noindent
${}^{18}\F+\e\rarr{}^{18}\O$\par             
\f@
      F  = 1.052E-4
@
\smallskip\noindent
${}^{18}\Ne+\e\rarr{}^{18}\F$\par            
\f@
      F  = 0.4146
@
\vfill\eject\noindent
${}^{1}\h+\p\rarr\e^+\nu+{}^{2}\h$\par       
\f@
      IF (T9 .LE. 3.) THEN
         F = 4.01E-15*T9M23*EXP(-3.380/T913)
              * (1.+.123*T913+1.09*T923+.938*T9)
      ELSE
         F = 1.16E-15
      END IF
@
\smallskip\noindent
${}^{1}\h+\e+\p\rarr\nu+{}^{2}\h$\par        
\f@
      IF (T9 .LE. 3.) THEN
         F = 1.36E-20*T9M76*EXP(-3.380/T913)
              * (1.-.729*T913+9.82*T923)
      ELSE
         F = 7.38E-12
      END IF
@
\smallskip\noindent
${}^{3}\He+\e\rarr\nu+{}^{3}\h$\par          
\f@
      IF (T9 .LE. 3.) THEN
         F = 7.71E-12*T932*EXP(-.2158/T9)
              * (1.+6.48*T9+7.48*T9**2+2.91*T9**3)
      ELSE
         F = 6.20E-9
      END IF
@
\smallskip\noindent
${}^{3}\He+\p\rarr\e^+\nu+{}^4\He$\par       
\f@
      IF (T9 .LE. 3.) THEN
         F = 8.78E-13*T9M23*EXP(-6.141/T913)
      ELSE
         F = 5.97E-15
      END IF
@
\smallskip\noindent
${}^{7}\Be+\e\rarr\nu\gamma+{}^{7}\Li$\par   
\f@
      IF (T9 .LE. 3.) THEN
         F = 1.34E-10/T912 * (1.-.537*T913+3.86*T923
              + .0027*EXP(2.515E-3/T9)/T9)
      ELSE
         F = 6.39E-10
      END IF
@
\smallskip\noindent
${}^{6}\He\rarr\e+{}^{6}\Li$\par             
\f@
      F = 0.859
@
\smallskip\noindent
${}^{9}\Li\rarr\e+{}^{9}\Be$\par             
\f@
      F = 0.9846
@
\smallskip\noindent
${}^{9}\Li\rarr\e\n+2{}^4\He$\par            
\f@
      F = 2.9538
@
\smallskip\noindent
${}^{10}\Be (\p,\gamma) {}^{11}\B$\par       
\f@
      F = 2.45E+6*T9M23*EXP(-10.39/T913)
@
\smallskip\noindent
${}^{16}\O (\p,\gamma) {}^{17}\F$\par        
\f@
      F = 1.50E+8/(T923*(1.+2.13*(1.-EXP(-0.728*T923))))
           *EXP(-16.692/T913)
@
\smallskip\noindent
${}^{17}\O (\p,\gamma) {}^{18}\F$\par        
\f@
      T9A = T9/(1.+2.69*T9)
      F = 7.97E+7*T9A**(5./6.)*T9M32*EXP(-16.712/T9A**(1./3.))
           + 1.51E+8*T9M23*EXP(-16.712/T913)*(1.+0.025*T913
           - 0.051*T923-8.82E-3*T9) + 1.56E+5*EXP(-6.272/T9)/T9
           + 1.31*T9M32*EXP(-1.961/T9)
@
\smallskip\noindent
${}^{4}\He (\n\n,\gamma) {}^{6}\He$\par      
\f@
      F = 4.04E-12/T9**2 * EXP(-9.585/T9)*(1.+.138*T9)
@
\smallskip\noindent
${}^{7}\Li (\n,\gamma) {}^8\Li$\par          
\f@
      F = 3.144E+3 + 4.26E+3*T9M32*EXP(-2.576/T9)
@
\smallskip\noindent
${}^{8}\Li (\n,\gamma) {}^{9}\Li$\par        
\f@
      F = 4.294E+4 + 6.047E+4*T9M32*EXP(-2.866/T9)
@
\smallskip\noindent
${}^{9}\Be (\n,\gamma) {}^{10}\Be$\par       
\f@
      F = 1.26E+3
@
\smallskip\noindent
${}^{10}\Be (\n,\gamma) {}^{11}\Be$\par      
\f@
      F = 1.32
@
\smallskip\noindent
${}^{11}\B (\n,\gamma) {}^{12}\B$\par        
\f@
      F = 7.29E+2+T9M32*(2.25E+3*EXP(-0.221/T9)
           +3.26E+4*EXP(-4.514/T9)+1.96E+4*EXP(-10.804/T9)
           +3.90E+4*EXP(-13.323/T9)+5.86E+4*EXP(-18.916/T9))
@
\smallskip\noindent
${}^{16}\O (\n,\gamma) {}^{17}\O$\par        
\f@
      F = 2.36E+1*(1.+4.45*T9)+9.66E+4*T9M32*EXP(-4.75/T9)
@
\smallskip\noindent
${}^{17}\O (\n,\gamma) {}^{18}\O$\par        
\f@
      F = 3.11*(1.+100.*T9)
@
\smallskip\noindent
${}^{14}\N (\n,\p) {}^{14}C$\par             
\f@
      F = 2.39E+5*(1.+.361*T912+.502*T9)
           + 1.112E+8*EXP(-4.983/T9)/T912
@
\smallskip\noindent
${}^{14}\O (\n,\p) {}^{14}\N$\par            
\f@
      F = 2.02E+8*(1.+.658*T912+.379*T9)
@
\smallskip\noindent
${}^{17}\F (\n,\p) {}^{17}\O$\par            
\f@
      F = 1.80E+8
@
\smallskip\noindent
${}^{18}\F (\n,\p) {}^{18}\O$\par            
\f@
      F = 1.80E+8
@
\smallskip\noindent
${}^{11}\Be (\p,\n) {}^{11}\B$\par           
\f@
      F = 1.71E+11*T9M23*EXP(-10.42/T913)
@
\smallskip\noindent
${}^{12}\C (\alpha,\gamma) {}^{16}\O$\par    
\f@
      F = 1.04E+8*EXP(-32.120/T913-(T9/3.496)**2)
           / (T9*(1.+.0489*T9M23))**2
           + 1.76E+8*EXP(-32.120/T913)/(T9*(1.+.2654*T9M23))**2
           + 1.25E+3*T9M32*EXP(-27.499/T9)
           + 1.43E-2*T9**5*EXP(-15.541/T9)
@
\smallskip\noindent
${}^{14}\C (\alpha,\gamma) {}^{18}\O$\par    
\f@
      F = 3.375E+8*EXP(-32.513/T913)/T9**2
           + 1.528E+9*T9M23*EXP(-32.513/T913-(T9/2.662)**2)
           * (1.+0.0128*T913-0.869*T923
           - 0.0779*T9+0.321*T943+0.0732*T953)
           + 9.29E-8*T9M32*EXP(-2.048/T9)
           + 2.77E+3*EXP(-9.876/T9)/T9**(4./5.)
@
\smallskip\noindent
${}^{14}\N (\alpha,\gamma) {}^{18}\F$\par    
\f@
      F = 7.78E+9*T9M23*EXP(-36.031/T913-(T9/0.881)**2)
           * (1.+0.012*T913+1.45*T923+0.117*T9+1.97*T943+0.406*T953)
           + 2.36E-10*T9M32*EXP(-2.798/T9)+2.03*T9M32*EXP(-5.054/T9)
           + 1.15E+4*T9M32*EXP(-12.310/T9)
@
\smallskip\noindent
${}^{14}\O (\alpha,\gamma) {}^{18}\Ne$\par   
\f@
      F = 9.47E+8*T9M23*EXP(-39.388/T913-(T9/.717)**2)
           * (1.+.011*T913+1.974*T923+.146*T9+3.036*T943+.572*T953)
           + 1.16E-1*T9M32*EXP(-11.733/T9)
           + 3.39E+1*T9M32*EXP(-22.609/T9)
           + 9.10E-3*T9**5*EXP(-12.159/T9)
@
\smallskip\noindent
${}^{11}\B (\alpha,\p) {}^{14}\C$\par        
\f@
      F = 8.403E+15*(1.+0.022*T913+5.712*T923+0.642*T9
           +15.982*T943+4.062*T953)
           *EXP(-31.914/T913-(T9/0.3432)**2)
           +4.944E+6*T9**(3./5.)*EXP(-11.26/T9)+T9M32
           *(5.44E-3*EXP(-2.868/T9)+2.419E+2*EXP(-5.147/T9)
           +4.899E+2*EXP(-5.157/T9))
@
\smallskip\noindent
${}^{11}\C  (\alpha,\p) {}^{14}\N$\par       
\f@
      T9A = T9/(1.+4.78E-2*T9+7.56E-3*T953/(1.+4.78E-2*T9)**(2./3.))
      F = 7.15E+15*T9A56*T9M32*EXP(-31.883/T9A13)
@
\smallskip\noindent
${}^{14}\O (\alpha,\p) {}^{17}\F$\par        
\f@
      F = 1.68E+13*T9M23*EXP(-39.388/T913-(T9/0.717)**2)
           * (1.+0.011*T913+13.117*T923+0.971*T9+85.295*T943
           + 16.061*T953)
           + 3.31E+4*T9M32*EXP(-11.733/T9)+1.79E+7*T9M32
           * EXP(-22.609/T9)
           + 9.E+3*T9**(11./3.)*EXP(-12.517/T9)
@
\smallskip\noindent
${}^{7}\Li (\p,\alpha\gamma) {}^4\He$\par    
\f@
      T9A=T9/(1.+.759*T9)
      F = 1.096E+9*T9M23*EXP(-8.472/(T913))
           -4.830E+8*(T9A**(5./6.))*T9M32*EXP(-8.472/(T9A**.333333333))
           +1.06E+10*T9M32*EXP(-30.442/T9)
           +1.56E+5*T9M23*EXP(-8.472/T913-(T9/1.696)**2)
           *(1.+.049*T913+2.498*T923+.860*T9+3.518*T943+3.080*T953)
           +1.55E+6*T9M32*EXP(-4.478/T9)
@
\smallskip\noindent
${}^{10}\Be (\p,\alpha) {}^{7}\Li$\par       
\f@
      F = 2.45E+11*T9M32*EXP(-10.39/T913)
@
\smallskip\noindent
${}^{11}\Be (\p,\alpha) {}^8\Li$\par         
\f@
      F = 8.57E+10*T9M23*EXP(-10.42/T913)
@
\smallskip\noindent
${}^{11}\B (\p,\alpha) 2{}^4\He$\par         
\f@
      F = 2.20E+12*T9M23*EXP(-12.095/T913-(T9/1.644)**2)
           *(1.+.034*T913+.140*T923+.034*T9+.190*T943+.116*T953)
           +4.03E+6*EXP(-1.734/T9)*T9M32
           +6.73E+9*EXP(-6.262/T9)*T9M32
           +3.88E+9*EXP(-14.154/T9)/T9
@
\smallskip\noindent
${}^{17}\O (\p,\alpha) {}^{14}\N$\par        
\f@
      F = 1.53E+7*T9M23*EXP(-16.712/T913-(T9/0.565)**2)
           * (1.+0.025*T913+5.39*T923+0.940*T9+13.5*T943+5.98*T953)
           + 2.92E+6*T9*EXP(-4.247/T9)
           + 0.1*(4.81E+10*T9*EXP(-16.712/T913-(T9/0.04)**2)
           + 5.05E-5*T9M32*EXP(-0.723/T9)
           + 1.31E+1*T9M32*EXP(-1.961/T9))
@
\smallskip\noindent
${}^{18}\O (\p,\alpha) {}^{15}\N$\par        
\f@
      F = 3.63E+11*T9M23*EXP(-16.729/T913-(T9/1.361)**2)
           * (1.+0.025*T913+1.88*T923+0.327*T9+4.66*T943+2.06*T953)
           + 9.9E-14*T9M32*EXP(-0.231/T9)+2.66E+4*T9M32*EXP(-1.67/T9)
           +2.41E+9*T9M32*EXP(-7.638/T9)+1.46E+9*EXP(-8.31/T9)/T9
@
\smallskip\noindent
${}^{18}\F (\p,\alpha) {}^{15}\O$\par        
\f@
      F = 9.52E+12*T9M23*EXP(-18.1/T913)
@
\vfill\eject\noindent
${}^{8}\Li (\alpha,\n) {}^{11}\B$\par        
\f@
      F = T9M32*(5.505E+6*EXP(-4.410/T9)
           +4.596E+8*EXP(-6.847/T9))
           +1.E+13*T9M23*EXP(-19.45/T913)*(2.02*T913
           +17.71*T923+17.65*T9+3.57*T943)
@
\smallskip\noindent
${}^{10}\Be (\alpha,\n) {}^{13}\C$\par       
\f@
      F = 3.64E+14*T9M23*EXP(-24.12/T913)
@
\smallskip\noindent
${}^{11}\Be (\alpha,\n) {}^{14}\C$\par       
\f@
      F = 4.51E+14*T9M23*EXP(-24.33/T913)
@
\smallskip\noindent
${}^{10}\B (\alpha,\n) {}^{13}\N$\par        
\f@
      F = 1.20E+13*T9M23*EXP(-27.989/T913-(T9/9.589)**2)
@
\smallskip\noindent
${}^{11}\B (\alpha,\n) {}^{14}\N$\par        
\f@
      F = 2.468E+15*(1.+7.519*T913+1.361*T923-14.972*T9
           -11.61*T943+18.145*T953)*EXP(-18.145/T913
           -(T9/0.7207)**2)
           +1.459E+7*T9**(3./5.)*EXP(-11.26/T9)+T9M32
           *(1.79*EXP(-2.868/T9)+1.678E+3*EXP(-5.147/T9)
           +5.358E+3*EXP(-5.157/T9))
@
\smallskip\noindent
${}^{13}\C (\alpha,\n) {}^{16}\O$\par        
\f@
      F = 6.77E+15*T9M23*EXP(-32.329/T913-(T9/1.284)**2)
           * (1.+.013*T913+2.04*T923+.184*T9)
           + 3.82E+5*T9M32*EXP(-9.373/T9)
           + 1.41E+6*T9M32*EXP(-11.873/T9)
           + 2.00E+9*T9M32*EXP(-20.409/T9)
           + 2.92E+9*T9M32*EXP(-29.283/T9)
@
\smallskip\noindent
${}^{17}\O (\n,\alpha) {}^{14}\C$\par        
\f@
      F = 3.11E+4*(1.+100.*T9)+2.12E+16*T9M23
           * EXP(-32.51/T913+21.11/T9-(2.33/T9)**2.51)/2.03
@
\smallskip\noindent
${}^{17}\F (\n,\alpha) {}^{14}\N$\par        
\f@
      F = 7.76E+9*(1.-1.15*T912+0.365*T9)*EXP(-(T9/2.798)**2)
           + 4.85E+10*T9M32*EXP(-15.766/T9)
@
\smallskip\noindent
${}^{18}\F (\n,\alpha) {}^{15}\N$\par        
\f@
      F = 6.28E+8*(1.-0.641*T912+0.108*T9)
@
\smallskip\noindent
${}^{2}\h (\d,\gamma) {}^4\He$\par           
\f@
      F = 4.84E+1*T9M23*EXP(-4.258/T913)
           * (1.+.098*T913-.203*T923-.139*T9+.106*T943+.185*T953)
@
\vfill\eject\noindent
${}^{6}\Li (\d,\n) {}^{7}\Be$\par            
\f@
      F = 1.48E+12*T9M23*EXP(-10.135/T913)
@
\smallskip\noindent
${}^{8}\Li (\d,\n) {}^{9}\Be$\par            
\f@
      F = 3.22E+11*T9M23*EXP(-10.357/T913)
@
\smallskip\noindent
${}^{14}\C (\d,\n) {}^{15}\N$\par            
\f@
      F = 4.27E+13*T9M23*EXP(-16.939/T913)
@
\smallskip\noindent
${}^{6}\Li (\d,\p) {}^{7}\Li$\par            
\f@
      F = 1.48E+12*T9M23*EXP(-10.135/T913)
@
\smallskip\noindent
${}^{7}\Li (\d,\p) {}^8\Li$\par              
\f@
      F = 8.31E+8*T9M32*EXP(-6.998/T9)
@
\smallskip\noindent
${}^{9}\Be (\p,\d) 2{}^4\He$\par             
\f@
      F = 2.11E+11*T9M23*EXP(-10.359/T913 - (T9/.520)**2)
           *(1.+.040*T913+1.09*T923+.307*T9+3.21*T943+2.30*T953)
           + 5.79E+8*EXP(-3.046/T9)/T9
           + 8.50E+8*EXP(-5.800/T9)/T934
@
\smallskip\noindent
${}^{3}\h (\t,2\n) {}^4\He$\par              
\f@
      F = 1.67E+9*T9M23*EXP(-4.872/T913)
           * (1.+.086*T913-.455*T923-.272*T9+.148*T943+.225*T953)
@
\smallskip\noindent
${}^{7}\Li (\t,\n\n\alpha) {}^4\He$\par      
\f@
      F = 8.81E+11*T9M23*EXP(-11.333/T913)
@
\smallskip\noindent
${}^{7}\Li (\t,\n) {}^{9}\Be$\par            
\f@
      F = 1.46E+11*T9M23*EXP(-11.333/T913)
@
\smallskip\noindent
${}^{9}\Be (\t,\n) {}^{11}\B$\par            
\f@
      F = 3.80E+12*T9M23*EXP(-14.02/T913)
           + 1.25E+8*T9M32*EXP(-4.43/T9)
@
\smallskip\noindent
${}^{3}\He (\t,\d) {}^4\He$\par              
\f@
      T9A = T9 / (1.+.128*T9)
      F = 5.46E+9*T9A56*T9M32*EXP(-7.733/T9A13)
@
\smallskip\noindent
${}^{1}\h (\p\n,\p) {}^{2}\h$\par            
\f@
      F = 1.42E-2*T9M32*EXP(-3.720/T913)
           * (1.+.784*T913+.346*T923+.690*T9)
@
\smallskip\noindent
${}^{7}\Be ({}^3\He,\p\p\alpha) {}^4\He$\par 
\f@
      F = 6.11E+13*T9M23*EXP(-21.793/T913)
@
\vfill\eject\noindent
${}^{14}\C (\n,\gamma) {}^{15}\C$\par        
\f@
      F = 1.08E+8*3.0E-5*T9
@
\smallskip\noindent
${}^{13}\N (\alpha,\p) {}^{16}\O$\par        
\f@
      T9A = T9/(1.+7.76E-2*T9+2.64E-2*T953/(1.+7.76E-2*T9)**(2./3.))
      F = 3.23E+17*T9A56*T9M32*EXP(-35.829/T9A13)
@
\smallskip\noindent
${}^{4}\He (2\alpha,\gamma) {}^{12}\C$\par   
\f@
      F = 2.79E-8*T9M32*T9M32*EXP(-4.4027/T9)
@
\smallskip\noindent
${}^{8}\Li (\p,\n\alpha) {}^{4}\He$\par      
\f@
      F = 1.031E+10*T9M23*EXP(-8.429/T913)
           +6.79E+5*T9M32*EXP(-1.02/T9)
           +3.28E+8*T9M32*EXP(-7.024/T9)
           +1.13E+9*T9**(-0.433)*EXP(-3.982/T9)
@
\smallskip\noindent
${}^{15}\N (\p,\alpha) {}^{12}\C$\par        
\f@
      F = 1.08E+12*T9M23*EXP(-15.251/T913-(T9/.522)**2)
           *(1.+.027*T913+2.62*T923+.501*T9+5.36*T943+2.60*T953)
           +1.19E+8*T9M32*EXP(-3.676/T9)
           +5.41E+8*EXP(-8.926/T9)/T912
           +4.72E+7*T9M32*EXP(-7.721/T9)
           +2.20E+8*T9M32*EXP(-11.418/T9)
@
\smallskip\noindent
${}^{18}\O (\n,\gamma) {}^{19}\O$\par        
\f@
      F = 21.2
@
\smallskip\noindent
${}^{20}\O \rarr \e+\bar\nu+{}^{20}\F$\par   
\f@
      F = 0.737
@
\smallskip\noindent
${}^{21}\F \rarr \e+\bar\nu+{}^{21}\Ne$\par  
\f@
      F = 0.234
@
\bigskip

The remaining reactions are estimated as follows.

\smallskip\noindent
${}^{8}\Li (\p,\gamma) {}^{9}\Be$\par        
\f@
      F = 6.27E+8*T9M23*EXP(-8.5/T913)*(1.+0.049*T913)
@
\smallskip\noindent
${}^{9}\Li (\p,\alpha) {}^{6}\He$\par        
\f@
      F = 1.03E+11*T9M23*EXP(-8.533*T9M13)
@
\smallskip\noindent
${}^{9}\Li (\d,\n) {}^{10}\Be$\par           
\f@
      F = 2.86E+11*T9M23*EXP(-10.41*T913)
@
\smallskip\noindent
${}^{12}\N (\n,\d) {}^{11}\C$\par            
\f@
      F = 1.32E+5
@
\smallskip\noindent
${}^{10}\Be (\alpha,\gamma) {}^{14}\C$\par   
\f@
      F = 5.82E+14*T9M23*EX(-24.12/T913)
           + 8.30E+7*T9M32*EX(-1.161/T9)
           + 1.01E+8*T9M32*EX(-6.731/T9)
@
\smallskip\noindent
${}^{11}\B (\alpha,\gamma) {}^{15}\N$\par    
\f@
      F = 4.3E+15*T9M23*EXP(-6.082/T913)
           + 6.8E+7*T9M32*EXP(-1.242/T9)
           + 7.5E+7*T9M32*EXP(-2.832/T9)
@
\smallskip\noindent
${}^{12}\N (\n,\p) {}^{12}\C$\par            
\f@
      F = 1.0E+12
@
\smallskip\noindent
${}^{8}\Li (\d,\p) {}^{9}\Li$\par            
\f@
      F = 2.58E+11*T9M23*EX(-10.34/T913)
          + 1.24E+8*T9M32*EXP(-2.95/T9)
@
\smallskip\noindent
${}^{9}\Li (\p,\n) {}^{9}\Be$\par            
\f@
      F = 1.03E+11*T9M23*EXP(-8.533/T913)
@
\smallskip\noindent
${}^{9}\Li (\alpha,\n) {}^{12}\B$\par        
\f@
      F = 8.82E+15*T9M23*EXP(-19.70/T913)
@
\smallskip\noindent
${}^{9}\Li (\p,\gamma) {}^{10}\Be$\par       
\f@
      F = 1.03E+11*T9M23*EXP(-8.533/T913)
           + 3.1E+5*T9M32*EXP(-11.61/T9)
@
\smallskip\noindent
${}^{13}\N (\n,\gamma) {}^{14}\N$\par        
\f@
      F = 1.32E+5
           + 1.25E+6*T9M23*EXP(-0.16/T9)
           + 1.32E+7*T9M23*EXP(-1.39/T9)
@
\smallskip\noindent
${}^{10}\B (\alpha,\gamma) {}^{14}\N$\par    
\f@
      F = 5.82E+14*T9M23*EXP(-27.98/T913)
           + 8.22E+7 *T9M32*EXP(-1.718/T9)
           + 2.99E+10*T9M32*EXP(-1.485/T9)
@
\smallskip\noindent
${}^{11}\C (\alpha,\gamma) {}^{15}\O$\par    
\f@
      F = 2.05E+16*T9M23*EXP(-31.88/T913)
           + 1.27E+8*T9M32*EXP(-0.928/T9)
           + 1.34E+8*T9M32*EXP(-3.017/T9)
           + 1.35E+8*T9M32*EXP(-3.365/T9)
@
\smallskip\noindent
${}^{8}\B (\alpha,\gamma) {}^{12}\N$\par     
\f@
      F = 8.67E+8*T9M23*EXP(-8.08/T913)
           * (1.+0.052*T913-0.448*T923-0.165*T9+0.144*T943+0.134*T953)
@
\smallskip\noindent
${}^{19}\O (\n,\gamma) {}^{20}\O$\par         
\f@
      F = 1.3E+5
@
\smallskip\noindent
${}^{17}\N (\alpha,\p) {}^{20}\O$\par         
\f@
      F = 3.3E+14*T9M23*EXP(-36.51/T913)
@
\smallskip\noindent
${}^{16}\C (\alpha,\gamma) {}^{20}\O$\par     
\f@
      F = 5.6E+16*T9M13*EXP(-32.82/T913)
@
\smallskip\noindent
${}^{20}\F (\n,\gamma) {}^{21}\F$\par         
\f@
      F = 1.3E+5
@
\smallskip\noindent
${}^{21}\F (\p,\alpha) {}^{18}\O$\par         
\f@
      F = 1.6E+13*T9M23*EXP(-18.10/T9)
@
\smallskip\noindent
${}^{17}\N (\alpha,\gamma) {}^{21}\F$\par     
\f@
      F = 3.3E+14*T9M23*EXP(-36.51/T913)
@

\vfill\eject

Table 1. Reactions changed since Walker et al (1991).
\bigskip
\centerline{
\vbox
  {\tabskip=0pt \offinterlineskip
   \def\tablerule{\noalign{\hrule}}
   \halign to351pt
      {\strut#&                                
       \vrule#\tabskip=2em plus2em&            %
       \hfil#\hfil&                            
       #\tabskip=2em plus 2 em&                %
       #\hfil&                                 
       #\tabskip=2em plus 2 em&                %
       #\hfil&                                 
       \vrule#\tabskip=0pt\cr                  %
       \tablerule
       &&\multispan5&\cr
       \strut&&\omit\hidewidth Reaction\hidewidth&&
                         \omit\hidewidth Notes\hidewidth&&
                         \omit\hidewidth Refs\hidewidth&\cr
       &&\multispan5&\cr
       \tablerule
       &&\multispan5&\cr
&& ${}^{3}\h+\e\rarr{}^{3}\He$                  && U    && T85          &\cr
&& ${}^{8}\Li\rarr\e+2{}^{4}\He$                && U    && T85          &\cr
&& ${}^{11}\Be\rarr\e+{}^{11}\B$                && A    && T85          &\cr
&& ${}^{8}\B+\e\rarr2{}^4\He$                   && U    && T85          &\cr
&& ${}^{12}\B\rarr\e+{}^{12}\C$                 && U    && T85          &\cr
&& ${}^{11}\C+\e\rarr{}^{11}\B$                 && U    && T85          &\cr
&& ${}^{14}\O+\e\rarr {}^{14}\N$                && U    && T85          &\cr
&& ${}^{15}\O+\e\rarr {}^{15}\N$                && U    && T85          &\cr
&& ${}^{17}\F+\e\rarr{}^{17}\O$                 && A    && T85          &\cr
&& ${}^{18}\F+\e\rarr{}^{18}\O$                 && A    && T85          &\cr
&& ${}^{18}\Ne+\e\rarr{}^{18}\F$                && A    && T85          &\cr
&& ${}^{1}\h+\p\rarr\e^+\nu+{}^{2}\h$           && A    && CF88         &\cr
&& ${}^{1}\h+\e+\p\rarr\nu+{}^{2}\h$            && A    && CF88         &\cr
&& ${}^{3}\He+\e\rarr\nu+{}^{3}\h$              && A    && CF88         &\cr
&& ${}^{3}\He+\p\rarr\e^+\nu+{}^4\He$           && A    && CF88         &\cr
&& ${}^{7}\Be+\e\rarr\nu\gamma+{}^{7}\Li$       && A    && CF88         &\cr
&& ${}^{6}\He\rarr\e+{}^{6}\Li$                 && A    && T85          &\cr
&& ${}^{9}\Li\rarr\e+{}^{9}\Be$                 && A    && LS78         &\cr
&& ${}^{9}\Li\rarr\e\n+2{}^4\He$                && A    && LS78         &\cr
&& ${}^{10}\Be (\p,\gamma) {}^{11}\B$           && A    && W69          &\cr
&& ${}^{16}\O (\p,\gamma) {}^{17}\F$            && A    && CF88         &\cr
&& ${}^{17}\O (\p,\gamma) {}^{18}\F$            && A    && CF88         &\cr
&& ${}^{4}\He (\n\n,\gamma) {}^{6}\He$          && A    && CF88         &\cr
&& ${}^{7}\Li (\n,\gamma) {}^8\Li$              && U    && WSK89        &\cr
&& ${}^{8}\Li (\n,\gamma) {}^{9}\Li$            && A    && MF89         &\cr
&& ${}^{9}\Be (\n,\gamma) {}^{10}\Be$           && A    && W69          &\cr
&& ${}^{10}\Be (\n,\gamma) {}^{11}\Be$          && A    && W69          &\cr
&& ${}^{11}\B (\n,\gamma) {}^{12}\B$            && US   && MF89         &\cr
&& ${}^{16}\O (\n,\gamma) {}^{17}\O$            && A    && W69          &\cr
&& ${}^{17}\O (\n,\gamma) {}^{18}\O$            && A    && W69          &\cr
&& ${}^{14}\N (\n,\p) {}^{14}C$                 && U    && CF88         &\cr
&& ${}^{14}\O (\n,\p) {}^{14}\N$                && A    && CF88         &\cr
&& ${}^{17}\F (\n,\p) {}^{17}\O$                && A    && W69          &\cr
&& ${}^{18}\F (\n,\p) {}^{18}\O$                && A    && W69          &\cr
&& ${}^{11}\Be (\p,\n) {}^{11}\B$               && A    && W69          &\cr
       \tablerule
       \noalign{\smallskip}
      }
  }
}
\vfil\eject
Table 1 (continued). Reactions changed since Walker et al (1991).
\bigskip
\centerline{
\vbox
  {\tabskip=0pt \offinterlineskip
   \def\tablerule{\noalign{\hrule}}
   \halign to351pt
      {\strut#&                                
       \vrule#\tabskip=2em plus2em&            %
       \hfil#\hfil&                            
       #\tabskip=2em plus 2 em&                %
       #\hfil&                                 
       #\tabskip=2em plus 2 em&                %
       #\hfil&                                 
       \vrule#\tabskip=0pt\cr                  %
       \tablerule
       &&\multispan5&\cr
       \strut&&\omit\hidewidth Reaction\hidewidth&&
                         \omit\hidewidth Notes\hidewidth&&
                         \omit\hidewidth Refs\hidewidth&\cr
       &&\multispan5&\cr
       \tablerule
       &&\multispan5&\cr
&& ${}^{12}\C (\alpha,\gamma) {}^{16}\O$        && U    && CF88         &\cr
&& ${}^{14}\C (\alpha,\gamma) {}^{18}\O$        && A    && CF88         &\cr
&& ${}^{14}\N (\alpha,\gamma) {}^{18}\F$        && A    && CF88         &\cr
&& ${}^{14}\O (\alpha,\gamma) {}^{18}\Ne$       && A    && CF88         &\cr
&& ${}^{11}\B (\alpha,\p) {}^{14}\C$            && US   && CF88         &\cr
&& ${}^{11}\C  (\alpha,\p) {}^{14}\N$           && U    && CF88         &\cr
&& ${}^{14}\O (\alpha,\p) {}^{17}\F$            && A    && CF88         &\cr
&& ${}^{7}\Li (\p,\alpha\gamma) {}^4\He$        && A    && CF88         &\cr
&& ${}^{10}\Be (\p,\alpha) {}^{7}\Li$           && A    && W69          &\cr
&& ${}^{11}\Be (\p,\alpha) {}^8\Li$             && A    && W69          &\cr
&& ${}^{11}\B (\p,\alpha) 2{}^4\He$             && U    && CF88         &\cr
&& ${}^{17}\O (\p,\alpha) {}^{14}\N$            && A    && CF88         &\cr
&& ${}^{18}\O (\p,\alpha) {}^{15}\N$            && A    && CF88         &\cr
&& ${}^{18}\F (\p,\alpha) {}^{15}\O$            && A    && W69          &\cr
&& ${}^{8}\Li (\alpha,\n) {}^{11}\B$            && US   && MF89         &\cr
&& ${}^{10}\Be (\alpha,\n) {}^{13}\C$           && A    && W69          &\cr
&& ${}^{11}\Be (\alpha,\n) {}^{14}\C$           && A    && W69          &\cr
&& ${}^{10}\B (\alpha,\n) {}^{13}\N$            && U    && CF88         &\cr
&& ${}^{11}\B (\alpha,\n) {}^{14}\N$            && US   && CF88         &\cr
&& ${}^{13}\C (\alpha,\n) {}^{16}\O$            && U    && CF88         &\cr
&& ${}^{17}\O (\n,\alpha) {}^{14}\C$            && A    && W69          &\cr
&& ${}^{17}\F (\n,\alpha) {}^{14}\N$            && A    && CF88         &\cr
&& ${}^{18}\F (\n,\alpha) {}^{15}\N$            && A    && CF88         &\cr
&& ${}^{2}\h (\d,\gamma) {}^4\He$               && A    && CF88         &\cr
&& ${}^{6}\Li (\d,\n) {}^{7}\Be$                && A    && MF89         &\cr
&& ${}^{8}\Li (\d,\n) {}^{9}\Be$                && A    && MF89         &\cr
&& ${}^{14}\C (\d,\n) {}^{15}\N$                && A    && KFKM91       &\cr
&& ${}^{6}\Li (\d,\p) {}^{7}\Li$                && A    && MF89         &\cr
&& ${}^{7}\Li (\d,\p) {}^8\Li$                  && A    && MF89         &\cr
&& ${}^{9}\Be (\p,\d) 2{}^4\He$                 && U    && CF88         &\cr
&& ${}^{3}\h (\t,2\n) {}^4\He$                  && A    && CF88         &\cr
&& ${}^{7}\Li (\t,\n\n\alpha) {}^4\He$          && A    && CF88         &\cr
&& ${}^{7}\Li (\t,\n) {}^{9}\Be$                && AS   && MF89         &\cr
&& ${}^{9}\Be (\t,\n) {}^{11}\B$                && A    && BK89         &\cr
&& ${}^{3}\He (\t,\d) {}^4\He$                  && A    && CF88         &\cr
       \tablerule
       \noalign{\smallskip}
      }
  }
}
\vfil\eject
Table 1 (continued). Reactions changed since Walker et al (1991).
\bigskip
\centerline{
\vbox
  {\tabskip=0pt \offinterlineskip
   \def\tablerule{\noalign{\hrule}}
   \halign to351pt
      {\strut#&                                
       \vrule#\tabskip=2em plus2em&            %
       \hfil#\hfil&                            
       #\tabskip=2em plus 2 em&                %
       #\hfil&                                 
       #\tabskip=2em plus 2 em&                %
       #\hfil&                                 
       \vrule#\tabskip=0pt\cr                  %
       \tablerule
       &&\multispan5&\cr
       \strut&&\omit\hidewidth Reaction\hidewidth&&
                         \omit\hidewidth Notes\hidewidth&&
                         \omit\hidewidth Refs\hidewidth&\cr
       &&\multispan5&\cr
       \tablerule
       &&\multispan5&\cr
&& ${}^{1}\h (\p\n,\p) {}^{2}\h$                && A    && CF88         &\cr
&& ${}^{7}\Be ({}^3\He,\p\p\alpha) {}^4\He$     && A    && CF88         &\cr
&& ${}^{14}\C (\n,\gamma) {}^{15}\C$            && A    && KFKM91       &\cr
&& ${}^{13}\N (\alpha,\p) {}^{16}\O$            && U    && CF88         &\cr
&& ${}^{4}\He (2\alpha,\gamma) {}^{12}\C$       && U    && CF88         &\cr
&& ${}^{8}\Li (\p,\n\alpha) {}^{4}\He$          && U    && BBL92        &\cr
&& ${}^{15}\N (\p,\alpha) {}^{12}\C$            && U    && CF88         &\cr
&& ${}^{18}\O (\n,\gamma) {}^{19}\O$            && A    && AS83         &\cr
&& ${}^{20}\O\rarr\e+\bar\nu+{}^{20}\F$         && A    && AS83         &\cr
&& ${}^{21}\F\rarr\e+\bar\nu+{}^{21}\Ne$        && A    && EV78         &\cr
&& ${}^{8}\Li (\p,\gamma) {}^{9}\Be$            && E    &&              &\cr
&& ${}^{9}\Li (\p,\alpha) {}^{6}\He$            && E    &&              &\cr
&& ${}^{9}\Li (\d,\n) {}^{10}\Be$               && E    &&              &\cr
&& ${}^{12}\N (\n,\d) {}^{11}\C$                && E    &&              &\cr
&& ${}^{10}\Be (\alpha,\gamma) {}^{14}\C$       && E    &&              &\cr
&& ${}^{11}\B (\alpha,\gamma) {}^{15}\N$        && E    &&              &\cr
&& ${}^{12}\N (\n,\p) {}^{12}\C$                && E    &&              &\cr
&& ${}^{8}\Li (\d,\p) {}^{9}\Li$                && E    &&              &\cr
&& ${}^{9}\Li (\p,\n) {}^{9}\Be$                && E    &&              &\cr
&& ${}^{9}\Li (\alpha,\n) {}^{12}\B$            && E    &&              &\cr
&& ${}^{9}\Li (\p,\gamma) {}^{10}\Be$           && E    &&              &\cr
&& ${}^{13}\N (\n,\gamma) {}^{14}\N$            && E    &&              &\cr
&& ${}^{10}\B (\alpha,\gamma) {}^{14}\N$        && E    &&              &\cr
&& ${}^{11}\C (\alpha,\gamma) {}^{15}\O$        && E    &&              &\cr
&& ${}^{8}\B (\alpha,\gamma) {}^{12}\N$         && E    &&              &\cr
&& ${}^{19}\O (\n,\gamma) {}^{20}\O$            && E    &&              &\cr
&& ${}^{17}\N (\alpha,\p) {}^{20}\O$            && E    &&              &\cr
&& ${}^{16}\C (\alpha,\gamma) {}^{20}\O$        && E    &&              &\cr
&& ${}^{20}\F (\n,\gamma) {}^{21}\F$            && E    &&              &\cr
&& ${}^{21}\F (\p,\alpha) {}^{18}\O$            && E    &&              &\cr
&& ${}^{17}\N (\alpha,\gamma) {}^{21}\F$        && E    &&              &\cr
       \tablerule
       \noalign{\smallskip}
      }
  }
}

In the notes column, U refers to a reaction whose rate has been
updated since Walker et al (1991); A, to a reaction which has
been added; S, to a reaction with several recent measurements;
and E, to a reaction which has been estimated.

The references give the most recent rate measurements
and refer to:
T85---Tuli (1985),
CF88---Caughlan and Fowler (1988),
LS78---Lederer and Shirley (1978),
EV78---Endt and Van der Leun (1978),
AS83---Ajzenberg-Selove (1983),
W69---Wagoner (1969),
WSK89---Wiescher et al (1989),
MF89---Malaney and Fowler (1989),
KFKM91---Kawano et al (1991),
BK89---Boyd and Kajino (1989),
BBL92---Becchetti et al (1992)

\vfil\eject

\centerline{References}

\refitem Ajzenberg-Selove, F.\ 1983, Nucl.\ Phys.\ A, 302, 1
\refitem Ajzenberg-Selove, F.\ 1985, Nucl.\ Phys.\ A, 433, 1
\refitem Ajzenberg-Selove, F.\ 1986, Nucl.\ Phys.\ A, 449, 1
\refitem Ajzenberg-Selove, F.\ 1988, Nucl.\ Phys.\ A, 490, 1
\refitem Alcock, C.R., Fuller, G.\ \& Mathews, G.J.\ 1987,
        ApJ, 320, 439
\refitem Andersen, J., Gustafsson, B.\ \& Lambert, D.L.,
        1984, AA 136, 65
\refitem Applegate, J.H., Hogan, C.J.\ \& Scherrer, R.J.,
        1988, ApJ, 329, 572
\refitem Barhoumi, S., Bogaert, G., Coc, A., Aguer, P., Kiener, J.,
        Lefebvre, A., Thibaud, J.-P., Baumann, H., Freiesleben, H.,
        Rolfs, C.\ \& Delbourgo-Salvador, P., 1991, preprint
\refitem Becchetti, F.D., Brown, J.A., Liu, W.Z., J\"anecke, J.W.,
        Roberts, D.A., Kolata, J.J., Smith, R.J., Lamkin, K.,
        Morsad, A., Warner, R.E., Boyd, R.N., \& Kalen, J.D.,
        1992, preprint
\refitem Boyd, R.N.\ \& Kajino, T., 1989, ApJ, 336, L55
\refitem Boyd, R.N., Kubono, S., Ikeda, N., Tanaka, M.H., Nomura, T.,
        Fuchi, Y., Kawashima, H., Ohura, M., Orihara, H., Yun, S.,
        Toyokawa, H., Yosoi, M., Ohnuma, H.,
        1992, preprint
\refitem Boyd, R.N., Tanihata, I., Inabe, N., Kubo, T., Nakagawa, T.,
        Suzuki, T., Yonokura, M., Bai, X.X., Kimura, K., Kubono, S.,
        Shimoura, S., Xu, H.S.\ \& Hirata, D., 1992, preprint
\refitem Brune, C.R., Kavanagh, R.W., Kellogg, S.E.\ \& Wang, T.R.,
        1991, Phys.\ Rev.\ C43, 875
\refitem Caughlan, G.R.\ \& Fowler, W.A.,
        1988, At.\ Dat.\ Nucl.\ Dat.\ Tabl., 40, 283
\refitem Delano, M.D., 1969, PhD thesis, New York University
\refitem Duncan, D.K., Lambert, D.L.\ \& Lemke, M., 1992, preprint
\refitem Endt, P.M.\ \& Van der Leun, C., 1978, Nucl.\ Phys.\ A, 310, 1
\refitem Fowler, W.A.\ \& Hoyle, F., 1964, ApJS, 9, 201
\refitem Gilmore, G., Edvardsson, B.\ \& Nissen, P.E.,
        1991, ApJ, 378, 17
\refitem Gilmore, G., Gustaffson, B., Edvardsson, B.\ \& Nissen, P.E.,
         1992, Nature, submitted
\refitem Hobbs, L.\ \& Pilachowski, C., 1988, ApJ, 326, L23
\refitem Hobbs, L.\ \& Thornburn, J., 1991, ApJ, in press
\refitem Kawano, L., 1992, preprint, FERMILAB-Pub-92/04-A
\refitem Kawano, L., Fowler, W.A., Kavanagh, R.W.\ \& Malaney, R.A.,
         1991, ApJ, 372, 1
\refitem Kawano, L., Schramm, D.N.\ \& Steigman, G.,
         1988, ApJ (Lett.), 327, 750
\refitem Krauss, L.L.\ \& Romanelli, P., 1990, ApJ, 358, 47
\refitem Kurki-Suonio, H., Matzner, R.A., Olive, K.A.\ \& Schramm, D.N.,
        1990, ApJ, 353, 406
\refitem Lederer, C.M., \& Shirley, V.S., 1978,
        Table of Isotopes (7th ed.\ Wiley, New York)
\refitem Malaney, R.A., \& Fowler, W.A.\ 1989, ApJ, 345, L5
\refitem Olive, K.A., \& Schramm, D.N.\ 1992, Nature (submitted)
\refitem Rebolo, R., Molaro, P., Abio, C., \& Beckman, J.E.,
        1988, AA, 193, 193
\refitem Ryan, S.G., Bessell, M.S., Sutherland, R.S., \& Norris, J.E.\ %
        1990, ApJ, 348, L57
\refitem Ryan, S.G., Norris, J.E., Bessell, M.S., \& Delyannis, C.P., 1992,
        ApJ (submitted)
\refitem Smith, M.S., Kawano, L.H., \& Malaney, R.A., preprint, OAP-716
\refitem Spite, J., \& Spite, F.\ 1982, A\&A, 115, 357
\refitem Steigman, G., Schramm, D.N., \& Gunn, J., 1977,
        Phys.\ Lett.\ 66B, 202
\refitem Steigman, G., \& Walker, T.P., 1992, ApJ, 385, L13
\refitem Terasawa, N., \& Sato, K., 1989, Prog.\ Theor.\ Phys., 1981, 1085
\refitem Terasawa, N., \& Sato, K., 1990, ApJ, 362, L47
\refitem Tuli, J.K.\ 1985, Nuclear Wallet Cards,
        (National Nuclear Data Center)
\refitem Turner, M.S., 1988, Phys.\ Rev.\ D37, 3049
\refitem Wagoner, R.V., 1969, ApJS, 18, 247
\refitem Walker, T.P., Steigman, G., Schramm, D.N., Olive, K.A., \&
        Fields, B., 1992, ApJ (submitted)
\refitem Walker, T.P., Steigman, G., Schramm, D.N., Olive, K.A., \&
        Kang, H.-S., 1991, ApJ, 376, 51
\refitem Wang, T.R., Vogelaar, R.B., \& Kavanagh, R.W., 1991,
        Phys.\ Rev.\ C43, 883
\refitem Wiescher, M., Steininger, R., \& K\"appeler, F., 1989,
        ApJ, 344, 464
\refitem Yang, J., Turner, M.S., Steigman, G., Schramm, D.N., \& Olive, K.A.,
        1984, ApJ, 281, 493

\vfil\eject

\centerline{Figure Captions}

\item{1.} The reaction network used in the code.  Estimated reactions are
        shown with dashed lines.
\item{2a.} Flow diagram for the standard model, with $\eta_{10}=3.0$
        (Flows to/from nuclides with $A\leq4$ are neglected).
\item{2b.} Flow diagram for the high n/p calculation, with n/p $=10$ and
        $\eta_{10}=3.0$ (Flows to/from nuclides with $A\leq4$ are neglected).
\item{3a.} \hefour\ mass fraction ($Y_p$) as a function of baryon to photon
        ratio ($\eta=n_b/n_\gamma$).  Neutron lifetime is 889.6 sec.
\item{3b.} Yields (number density relative to hydrogen) of \htwo, \hethree,
        \lisix\ and \liseven\ as functions of baryon to photon ratio
        ($\eta=n_b/n_\gamma$).  Neutron lifetime is 889.6 sec.
\item{3c.} Yields (number density relative to hydrogen) of \benine, \bten\ and
        \beleven\ as functions of baryon to photon ratio.  The bands for
        \benine\ and \bten\ are a result of the variation in the
        \liseven(t,n)\benine\ rate.  The maximum yields for \benine, \bten,
        \beleven\ within the range $2.8\le\eta_{10}\le4.0$ are
        $6\times10^{-18}$, $2\times10^{-19}$, $5\times10^{-17}$ respectively.
        Within $0.01\le\eta_{10}\le100$, maximum yields are
        $1\times10^{-14}$, $5\times10^{-19}$, $2\times10^{-14}$ respectively.
\item{4a.} \hefour\ mass fraction ($Y_p$) as a function of neutron to
        proton ratio, for $\eta_{10}=3.0$.
\item{4b.} Yields as a function of neutron to proton ratio,
        for $\eta_{10}=3.0$.

\vfill\eject

\end